\documentclass[twocolumn,showpacs,preprintnumbers,amsmath,amssymb]{revtex4}

\usepackage{graphicx}
\usepackage{dcolumn}
\usepackage{bm}

\newcommand{\bq}{\begin{equation}}
\newcommand{\eq}{\end{equation}}
\newcommand{\bqa}{\begin{eqnarray}}
\newcommand{\eqa}{\end{eqnarray}}
\newcommand{\nn}{\nonumber \\}

\def\be     {\begin{equation}}
\def\ee     {\end{equation}}
\def\bea        {\begin{eqnarray}}
\def\eea        {\end{eqnarray}}
\def\bnn    {\begin{eqnarray*}}
\def\enn    {\end{eqnarray*}}

\begin{document}

\title{Deconfinement in the presence of a Fermi surface}
\author{Ki-Seok Kim}
\affiliation{School of Physics, Korea Institute for Advanced
Study, Seoul 130-012, Korea}
\date{\today}

\begin{abstract}
U(1) gauge theory of non-relativistic fermions interacting via
compact U(1) gauge fields in the presence of a Fermi surface
appears as an effective field theory in low dimensional quantum
antiferromagnetism and heavy fermion liquids. We investigate
deconfinement of fermions near the Fermi surface in the effective
U(1) gauge theory. Our present analysis benchmarks the recent
investigation of quantum electrodynamics in two space and one time
dimensions ($QED_3$) by Hermele et al. [Phys. Rev. B {\bf 70},
214437 (2004)]. Utilizing a renormalization group analysis, we
show that the effective U(1) gauge theory with a Fermi surface has
a stable charged fixed point. Remarkably, the renormalization
group equation for an internal charge $e$ (the coupling strength
between non-relativistic fermions and U(1) gauge fields) reveals
that the conductivity $\sigma$ of fermions near the Fermi surface
plays the same role as the flavor number $N$ of massless Dirac
fermions in $QED_3$. This leads us to the conclusion that if the
conductivity of fermions is sufficiently large, instanton
excitations of U(1) gauge fields can be suppressed owing to
critical fluctuations of the non-relativistic fermions at the
charged fixed point. As a result a critical field theory of
non-relativistic fermions interacting via noncompact U(1) gauge
fields is obtained at the charged fixed point.
\end{abstract}


\maketitle

\section{Introduction}

Nature of quantum criticality is one of the central interests in
modern condensed matter physics. Especially, deconfined quantum
criticality has been proposed in various strongly correlated
electron systems such as low dimensional quantum
antiferromagnetism\cite{Laughlin_deconfinement,Senthil_deconfinement,
Kim1,Ichinose_deconfinement,Kleinert1,Review,Hermele_QED3,Kim2,Kleinert2,SL_Review,
Chubukov_SL,YBKim_SL,Fisher_SL,Motrunich_SL,SS_SL,Zhou_Wen} and
heavy fermion
liquids\cite{Senthil_Kondo,Coleman_Kondo,Pepin_Kondo,Kim_Kondo1,Kim_Kondo2}.
At these quantum critical points off-critical elementary degrees
of freedom such as magnons or electrons are proposed to break up
into more elementary particles with fractional quantum numbers. In
an opposite angle these off-critical elementary excitations can be
considered to be composites of the fractionalized elementary
excitations. One off-critical excitations in one phase can be
smoothly connected to the other off-critical ones in the other
phase via an appropriate fractionalization at the quantum critical
point. Physics of these quantum critical points is usually called
deconfined quantum criticality.

The main feature of deconfined quantum criticality is emergence of
gauge symmetries associated with fractionalized excitations.
Critical field theories describing deconfined quantum criticality
are naturally given by gauge theories, where fractionalized
elementary excitations interact via long range gauge fluctuations.
In the present paper we focus our attention on U(1) gauge theories
applicable to many proposed deconfined quantum critical points, as
will be discussed in the main section. Although U(1) gauge theory
formulation can explain non-Fermi liquid physics near quantum
critical points\cite{Review,SS_SL,Kim_Kondo1}, there exists one
fundamental difficulty originating from the fact that U(1) gauge
fields are basically compact. Compact U(1) gauge fields allow
instanton excitations expressing tunnelling events between
energetically degenerate but topologically inequivalent gauge
vacua. In the U(1) gauge theory instantons are nothing but
magnetic monopoles. It is well known that condensation of magnetic
monopoles results in confinement of fractionalized
excitations\cite{Polyakov}. If confinement arises from monopole
condensation, the effective U(1) gauge theory would be useless.
This is because the resulting field theory should be written in
terms of composites of fractionalized excitations. For the U(1)
gauge theory to be meaningful or physically working as a critical
field theory, the condensation of magnetic monopoles should be
forbidden.

Recently, it was shown that the condensation of magnetic monopoles
does not occur if the flavor number $N$ of massless Dirac fermions
is sufficiently large\cite{Hermele_QED3}. A relativistic U(1)
gauge theory in two space and one time dimensions usually dubbed
$(2+1)D$ quantum electrodynamics ($QED_{3}$) has a stable charged
fixed point in the limit of large
flavors\cite{Review,Hermele_QED3,Kim2,Kleinert2}. At the charged
fixed point critical fluctuations of Dirac fermions sufficiently
screen out the internal charge $e$ of Dirac fermions in the large
$N$ limit and thus, the corresponding magnetic charge $e_{m}$
becomes very large owing to the electro-magnetic duality $ee_{m} =
1$. Excitations of magnetic monopoles are suppressed. The
condensation of instantons (magnetic monopoles) is forbidden at
the stable charged fixed point owing to critical fluctuations of
Dirac fermions.

In the present paper we apply the analysis of the relativistic
U(1) gauge theory by Hermele et. al\cite{Hermele_QED3} to a
non-relativistic U(1) gauge theory with a Fermi surface. This U(1)
gauge theory has been considered to be a critical field theory in
the context of heavy fermion
liquids\cite{Senthil_Kondo,Coleman_Kondo,Pepin_Kondo,Kim_Kondo1,Kim_Kondo2}
and frustrated quantum
antiferromagnetism\cite{Motrunich_SL,SS_SL}. Utilizing a
renormalization group analysis, we show that the effective U(1)
gauge theory has a stable charged fixed point as the $QED_3$. In a
renormalization group equation for the internal charge $e$ of
non-relativistic fermions we find that {\it the conductivity
$\sigma$ of fermions plays the same role as the flavor number $N$
of Dirac fermions}. This leads us to the remarkable conclusion
that {\it if the conductivity is sufficiently large, instanton
excitations can be suppressed at the charged fixed point}. As a
result a critical field theory of non-relativistic fermions
interacting via noncompact U(1) gauge fields is obtained at the
charged critical point. In this critical field theory the coupling
strength between fermions and gauge fields is given by
$e/\sqrt{\sigma}$ as $e/\sqrt{N}$ in the $QED_3$. This implies
that at the charged fixed point correlation functions may be
systematically calculated in the $1/\sigma$ expansion as the $1/N$
expansion in the $QED_3$.

We would like to point out that a stable charged fixed point in
the non-relativistic U(1) gauge theory with a Fermi surface was
considered several years ago\cite{Polchinski,Altshuler,Gan}.
However, it should be noted that the previous
studies\cite{Polchinski,Altshuler} are verified by the $1/N$
expansion while the present study is justified by the $1/\sigma$
expansion.

\section{Review of $QED_3$}

In this section we review the relativistic U(1) gauge theory,
$QED_3$ in order to clarify the methodology utilized in the
non-relativistic U(1) gauge theory with a Fermi surface. We
consider the following effective $QED_3$ action \bqa && S =
\int{d^3x} \Bigl[ \bar{\psi}_{\alpha}\gamma_{\mu}(\partial_{\mu} -
ia_{\mu})\psi_{\alpha} + \frac{1}{2e^2}|\partial\times{a}|^2
\Bigr] . \eqa Here $\psi_{\alpha}$ is a massless Dirac spinor with
a flavor index $\alpha = 1, ..., N$ and $a_{\mu}$, a compact U(1)
gauge field. Eq. (1) was originally proposed to be an effective
action in one possible quantum disordered paramagnetism of SU(N)
quantum antiferromagnets on two dimensional square
lattices\cite{Review,Hermele_QED3,Kim2}. Utilizing the fermion
representation of the SU(N) antiferromagnetic Heisenberg model and
performing the Hubbard-Stratonovich transformation for appropriate
interaction channels, one can obtain an effective one body action
in terms of fermions and appropriate order parameter fields. In
this effective action a stable mean field phase is known to be the
$\pi$ flux state. In this mean field ground state low energy
elementary excitations are given by massless Dirac fermions near
nodal points showing gapless Dirac spectrum and U(1) gauge
fluctuations. This leads to Eq. (1) as a low energy effective
field theory in one possible quantum disordered paramagnetism of
the SU(N) Heisenberg model. Recently, Eq. (1) was also considered
to be an effective field theory in two dimensional geometrically
frustrated antiferromagnets\cite{SS_SL,Zhou_Wen}.

It is well known that the $QED_3$ with noncompact U(1) gauge
fields has a stable charged fixed
point\cite{Review,Hermele_QED3,Kim2,Kleinert2}. In order to see
this fixed point we introduce the relation of $e_{r}^{2} =
\Lambda{Z}_{a}e_{b}^{2}$, where $e_{r(b)}$ is the renormalized
(bare) internal charge of Dirac fermions and $Z_{a}$, the
renormalization constant of the gauge field $a_{\mu}$. This
relation shows how the internal gauge charge evolves by varying
the energy scale of our interest. The renormalization constant
$Z_{a}$ can be obtained from singular corrections to the
self-energy of the gauge field due to particle-hole polarizations
of massless Dirac fermions, given by $Z_{a} = 1 -
\lambda{N}{e}_{b}^{2}\ln\Lambda$. Here $\Lambda$ is a momentum
cut-off and $\lambda$, a positive numerical constant, where its
precise value is not important in the present consideration.
Inserting the expression of $Z_{a}$ into the relation of internal
gauge charges and performing the derivatives of
$\frac{de_{r}^2}{d\ln\Lambda} =
\frac{d\Lambda}{d\ln\Lambda}Z_{a}e_{b}^{2} +
\Lambda\frac{dZ_{a}}{d\ln\Lambda}e_{b}^{2}$, we obtain a
renormalization group equation for the internal charge
$e^{2}$\cite{Review,Hermele_QED3,Kim2,Kleinert2} \bqa
&&\frac{de^2}{d\ln\Lambda} = e^2 - \lambda{N}e^4 , \eqa where the
subscript $r$ in the renormalized charge $e_{r}$ is omitted. This
renormalization group equation expresses a change of the internal
charge $e^{2}$ as a function of the momentum cut-off $\Lambda$.
The first term represents a bare scaling dimension of $e^2$. In
$(2+1)D$ $e^2$ is relevant in contrast to the case of $(3+1)D$,
where it is marginal. The second term originates from the singular
correction to the self-energy of the U(1) gauge field by massless
Dirac fermions. This renormalization group equation leads us to a
stable charged fixed point of $e_{c}^{2} = {1}/{\lambda{N}}$ in
the $QED_3$.

A next question is if the charged fixed point remains stable after
admitting instanton excitations. Using the electromagnetic
duality, Hermele et. al obtained the following renormalization
group equations of magnetic charge $g = {1}/{e^2}$ and instanton
fugacity $y_m$\cite{Hermele_QED3}, \bqa && \frac{dg}{d\ln\Lambda}
= -g + \lambda{N} - \alpha{y_m^2}{g^3} , \nn &&
\frac{dy_m}{d\ln\Lambda} = (3 - \beta{g})y_{m}, \eqa where
$\alpha$ and $\beta$ are positive numerical constants. In the
absence of massless Dirac fermions ($N = 0$) Eq. (3) is reduced to
the standard renormalization group equation for the sine-Gordon
theory describing three dimensional Coulomb (monopole)
gas\cite{Polyakov}. The last term in the first equation results
from screening effects by monopole and anti-monopole pairs in the
sine-Gordon theory. On the other hand, the second term
$\lambda{N}$ is the contribution of massless Dirac fermions,
originating from Eq. (2) via the electromagnetic duality $g =
e^{-2}$. This term leads a magnetic charge to have a large fixed
point value proportional to $N$, i.e., $g_{c} = \lambda{N}$ in the
large $N$ limit. This large magnetic charge makes the instanton
fugacity $y_{m}$ go to zero at the charged fixed point. This is
the signal for suppression of instanton excitations. Although the
above renormalization group equations are approximate, there
exists a rather convincing argument\cite{Hermele_QED3,Conformal}.
An important basis for this argument is the existence of a charged
critical point. At the scale invariant fixed point it is shown
that the scaling dimension of an instanton insertion operator is
proportional to the order of $N$\cite{Conformal}. This leads one
to the conclusion that in the large $N$ limit the internal flux
changing operators are irrelevant at the critical fixed point,
indicating the suppression of instanton excitations.

A critical field theory in terms of massless Dirac fermions
interacting via noncompact U(1) gauge fields is obtained at the
charged fixed point \bqa && S_{c} = \int{d^3x} \Bigl[
\bar{\psi}_{\alpha}\gamma_{\mu}(\partial_{\mu} -
i\frac{e_{c}}{\sqrt{N}}a_{\mu})\psi_{\alpha} \nn && +
\frac{1}{16}(\partial\times{a})\frac{1}{\sqrt{-\partial^2}}(\partial\times{a})
\Bigr] . \eqa At the tree level it can be easily checked that this
effective action has scale invariance. Notice that the Maxwell
kinetic energy of the gauge field was ignored. Because the scaling
dimension of $|\partial\times{a}|^{2}$ is larger than $3$, the
Maxwell term is irrelevant at the charged critical point. The
non-Maxwell kinetic energy of the gauge field arises from the
contribution of critical Dirac fermions. We emphasize that
integration over the Dirac fermions should be understood in the
renormalization group sense. It is noted that the critical
coupling constant between the Dirac fermions and gauge fields is
given by $e_{c}/\sqrt{N}$ after replacing $a_{\mu}$ with
$a_{\mu}/\sqrt{N}$. Thus, correlation functions can be
systematically calculated in the $1/N$ expansion at the charged
fixed point. In the following we discuss that the non-relativistic
U(1) gauge theory with a Fermi surface has the similar structure
with the $QED_3$.

\section{Deconfinement in the presence of a Fermi surface}

\subsection{Effective Field Theory}

Now we consider the following U(1) gauge theory in terms of
non-relativistic fermions interacting via compact U(1) gauge
fields \bqa && S_{\chi} = \int{d^3x} \Bigl[
\chi_{\alpha}^{\dagger}(\partial_{\tau} - ia_{\tau} -
\mu)\chi_{\alpha} + \frac{1}{2m}|({\vec \nabla} - i{\vec
a})\chi_{\alpha}|^{2} \nn && +
\frac{1}{2e^2}|\partial\times{a}|^{2} \Bigr] . \eqa Here
$\chi_{\alpha}$ represents a fermion field with spin $\alpha =
\uparrow, \downarrow$ and $\mu$, its chemical potential. $a_{\mu}
= (a_{\tau}, {\vec a})$ is a compact U(1) gauge field. The Maxwell
kinetic energy of the gauge field can be considered to arise from
the contribution of high energy excitations. Eq. (5) is proposed
to be an effective field theory in various strongly correlated
electron systems such as low dimensional geometrically frustrated
quantum antiferromagnetism\cite{Motrunich_SL,SS_SL}, heavy fermion
liquids\cite{Senthil_Kondo,Kim_Kondo1,Kim_Kondo2} and strange
metals of high $T_c$ cuprates\cite{Polchinski,Altshuler,Nagaosa1}.
We briefly review how Eq. (5) appears to be an effective action in
the geometrically frustrated quantum antiferromagnets on two
dimensional triangular lattices\cite{Motrunich_SL,SS_SL}.
Utilizing the slave rotor representation of the Hubbard
model\cite{Rotor} and performing the Hubbard-Stratonovich
transformation for appropriate interaction channels, Lee and Lee
obtained an effective one body action in terms of fermionic
spinons and bosonic rotors coupled to hopping order
parameters\cite{SS_SL}. In the one body effective action the
authors found that near a metal-insulator transition a stable mean
field phase is a zero flux state. In this mean field ground state
low energy elementary excitations are given by gapless
non-relativistic spinons near a Fermi surface, gapped bosonic
rotors and compact U(1) gauge fields. Gapped bosons can be safely
integrated out to produce the Maxwell kinetic energy for the U(1)
gauge fields. As a result Eq. (5) is obtained to describe a spin
liquid phase near the Mott critical point.

In the present paper we assume that bosonic excitations are gapped
and thus, consider a spin liquid Mott insulator. This allows us to
investigate the fermion-only theory Eq. (5). The role of gapped
bosonic excitations is to generate the Maxwell kinetic energy for
the U(1) gauge field.

\subsection{Deconfinement in the presence of a Fermi surface}

Now we examine the deconfinement of non-relativistic fermions near
a Fermi surface. As mentioned in the introduction, our strategy is
basically the same as that of Hermele et al.\cite{Hermele_QED3}.
We first check whether there exists a stable charged fixed point
and then, investigate the stability of the charged fixed point
against instanton excitations. Before doing this, we linearize the
non-relativistic spectrum of $\chi_{\alpha}$ fermions near the
Fermi surface \bqa && S_{\chi} = \int {d^{3}x} \Bigl[
\chi^{\dagger}_{\alpha}\Bigl([\partial_{0} - ia_{0}] + {\vec
v}_{F}\cdot[i{\vec \nabla} + {\vec k}_{F} + {\vec a}]
\Bigr)\chi_{\alpha} \nn && + \frac{1}{2e^2}|\partial\times{a}|^{2}
\Bigr] , \eqa where $v_{F}$ is a Fermi velocity and $k_{F}$, a
Fermi wave vector. In the absence of long range gauge interactions
($e^2 = 0$) the resulting field theory describes noninteracting
fermions near the Fermi surface. This free fermion theory is a
trivial critical field theory at the Fermi liquid fixed
point\cite{RG}, more accurately, Fermi gas fixed point. The Fermi
liquid fixed point corresponds to the free Dirac fixed point in
the $QED_3$. As discussed in section II, the free Dirac fixed
point is unstable against long range U(1) gauge interactions. It
is naturally expected that the Fermi liquid fixed point is also
unstable against U(1) gauge fluctuations.

Just as the case of $QED_3$, we introduce the relation of
$e_{r}^{2} = \Lambda{Z}_{a}e_{b}^{2}$ between the renormalized and
bare internal charges, $e_{r}$ and $e_{b}$, respectively, where
$Z_{a}$ is the renormalization constant of the gauge field
$a_{\mu}$. Remember that singular corrections to the self-energy
of the gauge field due to particle-hole excitations of fermions
near the Fermi surface contribute to the renormalization constant
$Z_{a}$. Integrating over the fermions near the Fermi surface, we
obtain the following expression for an effective action, $S_{a} =
- Trln\Bigl([\partial_{0} - ia_{0}] + {\vec v}_{F}\cdot[i{\vec
\nabla} + {\vec k}_{F} + {\vec a}] \Bigr)$. Expanding the
logarithmic term to quadratic order for the U(1) gauge field
$a_{\mu}$, we obtain $S_{a} =
\frac{1}{2}\sum_{q,\omega}a_{\mu}(q,i\omega)\Pi_{\mu\nu}(q,i\omega)a_{\nu}(-q,-i\omega)$,
where $\Pi_{\mu\nu}(q,i\omega)$ is the density-density ($\mu = \nu
= \tau$) or current-current ($\mu, \nu = x, y$) correlation
function of gapless fermions. In this expression the time and
space components of the gauge field decouple. Since the time
component is screened by density fluctuations ($\Pi_{\tau\tau}$)
of gapless fermions and gives rise to only a short-range
interaction, it's sufficient to consider the spatial components
(labeled $i, j = x, y$) only\cite{Nagaosa1}. The current-current
correlation function is given by $\Pi_{ij}(x, \tau) = -
\langle{T}_{\tau}[J_{Fi}(x,\tau)J_{Fj}(0,0) -
\delta_{ij}\rho_{F}\delta(x)\delta(\tau)]\rangle$, where $J_{Fi} =
v_{Fi}\chi^{\dagger}_{\alpha}\chi_{\alpha}$ and $\rho_{F} =
\chi^{\dagger}_{\alpha}\chi_{\alpha}$ are the current and density
operators of fermions, respectively. It is convenient to choose
the Coulomb gauge ${\vec \nabla}\cdot{\vec a} = 0$, in which case
the spatial part of the gauge field is purely transverse. It
should be noted that since the density term in the current-current
correlation function originates from the $a_{i}^{2}$ term in Eq.
(5), it does not arise from Eq. (6) owing to the linearization of
a fermion dispersion near the Fermi surface. For the gauge field
to be transverse, this term should be taken into account
explicitly. It is well known that the transverse current-current
correlation function is given
by\cite{Polchinski,Altshuler,Gan,Nagaosa1,Tsvelik} \bqa &&
\Pi_{ij}(q, i\omega) = \Bigl(\delta_{ij} -
\frac{q_{i}q_{j}}{q^{2}}\Bigr)\Pi(q, i\omega) , \nn && \Pi(q,
i\omega) = \sigma|\omega| + \chi{q}^{2} , \eqa where $\sigma$ and
$\chi$ are the conductivity and diamagnetic susceptibility of
fermions near the Fermi surface. In appendix A we show this
derivation. As a result we obtain the effective gauge action in
the Coulomb gauge \bqa && S_{eff} =
\frac{1}{2}\sum_{q,\omega}[(\frac{1}{e^2} + \chi)q^{2} +
\sigma|\omega|]\Bigl(\delta_{ij} -
\frac{q_{i}q_{j}}{q^{2}}\Bigr)\nn&&~~~~~~~~~~a_{i}(q,i\omega)a_{j}(-q,-i\omega)
. \eqa The above expression can be easily expected. The
transverseness is naturally understood in the respect that U(1)
gauge symmetry restricts the resulting dynamics of gauge fields.
Since the fermions are gapless excitations, singular corrections
are expected to arise, renormalizing the internal charge $e$ as
the case of $QED_3$. In the non-Maxwell kinetic energy of the
gauge field a new feature is emergence of the conductivity
$\sigma$ of fermions. This is reflection of the Fermi surface.
Note that the non-Maxwell kinetic energy depends on the absolute
value of frequency. This indicates dissipative dynamics of the
gauge field. In the present paper we consider the case of Ohmic
dissipation, where the conductivity $\sigma$ is given by a
constant value depending on the density of states and mean free
time of fermions near the Fermi surface. We would like to comment
that in Eq. (8) the conductivity $\sigma$ lies in the same place
as the flavor number $N$ of Dirac fermions in the $QED_{3}$ Eq.
(4). This leads us to expect that the conductivity plays the same
role as the flavor number. If so, the expansion of the logarithmic
term to the second order for the U(1) gauge field can be justified
in the $1/\sigma$ expansion as the $1/N$ expansion in the $QED_3$.
The diamagnetic susceptibility is given by $\chi \sim m^{-1}$ in a
Fermi liquid with $m$, a mass of fermions\cite{Nagaosa1,Tsvelik}.

In order to justify Eq. (8) we expand the resulting logarithmic
term to higher order and write down the effective gauge action for
transverse gauge fields $a_{t}$ in a highly schematic form \bqa &&
S_{a} = \sigma\int{d^3q}|\omega||a_{t}|^{2} +
\int{(d^3q)^{3}}b_{4}f(q)|a_{t}|^{4} + O(a_{t}^{6}) , \nn \eqa
where $f(q)$ is a function of momentum and frequency. The
coefficient $b_{4}$ of nonlinear gauge interactions is given by
$b_{4} \sim
\int({d\Omega}{d^2k})G^{4}(k,\Omega)$\cite{Nonlinear_interactions1,Nonlinear_interactions2,Nagaosa_book}
with the single particle green function $G(k,\Omega) = [\Omega -
\epsilon_{k} + i\eta]^{-1}$ under $i\Omega \rightarrow \Omega +
i\eta$. The main point is whether the coefficient $b_{4}$ is
proportional to $\sigma^{2}$, square of the conductivity of
fermions near the Fermi surface. In this case the $1/\sigma$
expansion would be broken. If one utilizes the linearized spectrum
near the Fermi surface, i.e., $\epsilon_{k} = {\vec
v}_{F}\cdot({\vec k} - {\vec k}_{F})$, the integral over momentum
in the expression of $b_{4}$ would vanish owing to its multiple
pole
structure\cite{Nonlinear_interactions1,Nonlinear_interactions2}.
This implies that the integral contribution of $b_{4}$ results
from fermions with high energies, of the order of bandwidth, where
the spectrum cannot be
linearized\cite{Nonlinear_interactions1,Nonlinear_interactions2}.
This leads the coefficient $b_{4}$ to a constant
value\cite{Nonlinear_interactions1,Nonlinear_interactions2,Nagaosa_book}.
Other coefficients in higher order terms are also given by some
constants\cite{Nonlinear_interactions1}. Now we can see how the
$1/\sigma$ expansion works. For the gaussian gauge action to be
finite in the large $\sigma$ limit, fluctuations of gauge fields
should follow $a_{t} \sim 1/\sqrt{\sigma}$. Then, the nonlinear
terms are apparently higher order in the $1/\sigma$ expansion than
the leading gaussian term. This justifies the $1/\sigma$ expansion
for the non-Maxwell kinetic energy of the gauge field $a_{\mu}$. A
similar argument for the $1/N$ expansion in the $QED_3$ can be
found in Ref. \cite{Hermele_QED3}.

The $1/\sigma$ expansion may be understood {\it physically} in the
following way. In the $1/N$ expansion the flavor number $N$ of
Dirac fermions can be considered to be the number of screening
channels for gauge interactions. In a similar way the conductivity
is associated with the screening channels. In the case of Ohmic
dissipation the conductivity is given by $\sigma =
ne^{2}\tau_{tr}/m$, where $n$, $e$, $\tau_{tr}$, and $m$ are the
density, charge, transport time, and mass of fermions. It should
be noted that the density is involved with the conductivity. In
this respect the conductivity may be considered to be the number
of screening channels for gauge interactions.

The resulting non-Maxwell kinetic energy (Landau damping term) has
the same scaling as that in the $QED_3$ Eq. (4). This singular
correction in Eq. (8) leads to the following renormalization
constant $Z_{a} = 1 - \gamma\sigma{e}_{b}^{2}\ln\Lambda$, where
$\Lambda$ is a momentum cut-off and $\gamma$, a positive numerical
constant. Its precise value is not important in the present
consideration. Inserting this expression of $Z_{a}$ into the
relation of internal charges and performing derivatives with
respect to $\ln\Lambda$ as the case of $QED_3$, we reach a
renormalization group equation for the internal charge\cite{Wang}
\bqa && \frac{de^2}{d\ln\Lambda} = e^2 - \gamma\sigma{e}^{4} ,
\eqa where the subscript $r$ in the renormalized charge
$e_{r}^{2}$ is omitted. The first term represents a bare scaling
dimension of $e^2$ in $(2+1)D$, and the second term originates
from the singular correction to the self-energy of the U(1) gauge
field by non-relativistic fermions. Remarkably, this
renormalization group equation has the essentially same structure
as Eq. (2) in the $QED_3$ if the flavor number $N$ is replaced
with the conductivity $\sigma$. The Fermi liquid fixed point of
$e^{2} = 0$ is unstable against a nonzero value of internal
charge. A finite internal charge drives a renormalization group
flow away from the Fermi liquid fixed point, terminating at the
stable charged fixed point of $e_{c}^{2} = 1/(\gamma\sigma)$. The
effective U(1) gauge theory $S_{\chi}$ in Eq. (5) has a stable
charged fixed point as the $QED_3$ Eq. (1).

A next job is to examine the stability of the charged critical
point against instanton excitations. Using the electromagnetic
duality, we first obtain a renormalization group equation for
magnetic charge $g = {1}/{e^2}$ \bqa && \frac{dg}{d\ln\Lambda} = -
g + \gamma\sigma - \bar{\alpha}{y}_{m}^{2}{g}^{3} , \eqa where
$\bar{\alpha}$ is a positive numerical constant. The first and
second terms in right hand side originate from Eq. (10) via the
electromagnetic duality $g = e^{-2}$. The last term results from
the screening effect of monopole and anti-monopole pairs in a
non-relativistic sine-Gordon theory, $S_{sG} =
\int\frac{{d^2k}{d\omega}}{(2\pi)^{3}} \frac{1}{2g}(k^{2} +
\sigma^{-1}g|\omega|k^{2})|\varphi(k,\omega)|^{2} -
\int{d^2x}{d\tau} y_{m}\cos\varphi(x,\tau)$\cite{Nagaosa2,Herbut},
where $\varphi$ is a magnetic potential field mediating
interactions between magnetic monopoles and $y_{m}$, monopole
(instanton) fugacity. We note that owing to the dissipative
dynamics of the gauge field in Eq. (8) the above sine-Gordon
action has nontrivial momentum and frequency dependencies in the
kinetic energy of the $\varphi$ fields in contrast to the standard
sine-Gordon action, $S_{sG} =
\int\frac{{d^2k}{d\omega}}{(2\pi)^{3}} \frac{1}{2g}(k^{2} +
\omega^{2})|\varphi(k,\omega)|^{2} - \int{d^2x}{d\tau}
y_{m}\cos\varphi(x,\tau)$\cite{Polyakov}. The non-relativistic
sine-Gordon action leads to the following renormalization group
equation for the monopole fugacity $y_m$\cite{Nagaosa2,Herbut}
\bqa && \frac{dy_m}{d\ln\Lambda} = \Bigl(2 -
\bar{\beta}\sigma\ln(1 + \bar{\varrho}\sigma^{-1}g) \Bigr)y_{m} ,
\eqa where ${\bar \beta}$ and ${\bar \varrho}$ are positive
numerical constants. A detailed derivation of Eq. (12) can be
found in Eq. (B10) of Ref. \cite{Herbut}. Eq. (11) and Eq. (12)
yield that instanton (monopole) excitations can be suppressed at
the charged critical point in the large $\sigma$ limit. Eq. (11)
shows that the magnetic charge $g$ can have a large fixed point
value proportional to $\sigma$, i.e., $g_{c} = \gamma\sigma$ in
the large $\sigma$ limit. Inserting this fixed point value into
Eq. (12), one can easily find that the monopole fugacity goes to
zero in the large $\sigma$ limit. This is in contrast to the
result of Ref. \cite{Herbut}. The reason why there is no phase
transition in Ref. \cite{Herbut} lies in the introduction of a
${\hat a}_{\omega}$ parameter\cite{Herbut_RG}. However, the
presence of the ${\hat a}_{\omega}$ parameter destroys the charged
critical point even in the absence of instanton
excitations\cite{Herbut_RG}. In this respect we think that
introduction of the ${\hat a}_{\omega}$ parameter is not fully
justified. If this parameter is ignored, Kosterlitz-Thouless
($KT$) {\it like} phase transition is expected as a
confinement-deconfinement transition\cite{Herbut_RG}. This
possibility is distinct from our scenario since the structure of
Eq. (11) is essentially different from that of the renormalization
group equation in the $KT$ transition\cite{Kleinert1,Herbut_RG}.
Although the present result seems to be consistent with the
previous analytical study\cite{Nagaosa2} arguing the existence of
a finite critical conductivity for the confinement-deconfinement
transition, the nature of the transition would be different. We
would like to point out a report of Monte Carlo simulation
claiming deconfinement of non-relativistic
particles\cite{Numerical}. In the study the authors investigated
an effective nonlocal gaussian gauge action. From their Monte
Carlo simulation they argued that deconfinement of
non-relativistic particles always occurs. This is not contrast to
the present result in the sense that the present analysis can be
applied in the large $\sigma$ limit.

It should be noted that the above renormalization group equations,
Eq. (11) and Eq. (12) are approximate since they are obtained in
the gaussian approximation for the U(1) gauge fields. In order to
overcome this level of approximation it is necessary to apply the
methodology of Ref. \cite{Conformal} in the relativistic U(1)
gauge theory to the non-relativistic one. Remember that the
important basis of this nonperturbative argument is the existence
of a scale invariant critical point. In this respect we expect
that scaling dimensions of instanton insertion operators may be
given by the order of $\sigma$ as the order of $N$ in the $QED_3$.
This important issue should be addressed near future.

A critical field theory in terms of non-relativistic fermions
interacting via noncompact U(1) gauge fields is obtained at the
charged fixed point in the Coulomb gauge \bqa && S_{c} =
\int{d^3x} \Bigl[ \chi^{\dagger}_{\alpha}\Bigl(\partial_{0} +
{\vec v}_{F}\cdot[i{\vec \nabla} + {\vec k}_{F} +
\frac{e_{c}}{\sqrt{\sigma}}{\vec a}] \Bigr)\chi_{\alpha} \Bigr]
\nn && + \frac{1}{2}\sum_{q,\omega}\Bigl(\frac{1}{\sigma} +
\frac{|\omega|}{q^2}\Bigr) f_{xy}^{2}(q,\omega) , \eqa where the
field strength tensor $f_{xy}$ is given by $f_{xy} =
\partial_{x}a_{y} - \partial_{y}a_{x}$ in real space.
In the non-relativistic case the Maxwell kinetic energy of the
gauge field cannot be ignored since it is not irrelevant at the
charged critical point. If we assign the scaling dimensions of
${\vec a}$ and $\chi_{\alpha}$ as $[{\vec a}] = [{\vec q}]$ and
$[\chi_{\alpha}] = [{\vec q}]^{3/2}$ under $[\omega] = [{\vec
q}]^{2}$ with $[O]$, the scaling dimension of the variable $O$,
the above effective action has scale invariance at the tree level
except the time derivative term for the $\chi_{\alpha}$ fermions.
This can be resolved by the self-energy correction of fermions via
dissipative gauge interactions. Performing the standard one loop
calculation, we can easily find $[\Sigma] = [\omega]^{1/2}$, where
$\Sigma$ is the self-energy of the $\chi_{\alpha}$ fermions. Then,
the resulting effective action including the self-energy
correction has scaling invariance\cite{Polchinski,Altshuler}. We
note that the Maxwell kinetic energy is higher order than the
singular non-Maxwell kinetic energy in the $1/\sigma$ expansion.
Remember that nonlinear interactions between gauge fields are the
order of $1/\sigma^{2}$. In this respect it is consistent to keep
the Maxwell term in the $1/\sigma$ expansion. We would like to
point out that the critical coupling constant between the
non-relativistic fermions and gauge fields is given by
$e_{c}/\sqrt{\sigma}$ after replacing ${\vec a}$ with ${\vec
a}/\sqrt{\sigma}$. This clarifies the fact that the conductivity
$\sigma$ plays the same role as the flavor number $N$ of Dirac
fermions at the charged critical point. This implies that
correlation functions may be systematically evaluated in the
$1/\sigma$ expansion at the charged fixed point as the $1/N$
expansion in the $QED_3$.

\subsection{Discussion: Effect of Disorder on Deconfined Quantum Criticality}

In this section we discuss effects of nonmagnetic disorders on
deconfined fermions near the Fermi surface at the charged fixed
point. Recently, the role of nonmagnetic impurities in the
relativistic critical field theory Eq. (4) was investigated by the
present author\cite{Kim2}. In contrast to the free Dirac theory in
two space dimensions\cite{Lee_disorder,Fisher_disorder} long range
gauge interactions reduce strength of disorders and induce a
delocalized state at zero temperature\cite{Kim2}. The presence of
nonmagnetic disorders destabilizes the free Dirac fixed point. The
renormalization group flow goes away from the fixed point,
indicating localization\cite{Lee_disorder,Fisher_disorder}. On the
other hand, the charged fixed point in the $QED_3$ remains stable
at least against weak randomness\cite{Kim2}. A new unstable fixed
point separating delocalized and localized phases is
found\cite{Kim2}. The renormalization group flow shows that the
effect of random potentials vanishes if we start from sufficiently
weak disorders. In the present critical theory Eq. (13) a similar
result is expected. {\it Deconfined fermions near the Fermi
surface would remain delocalized at least against weak
randomness}. However, it should be considered that nonmagnetic
disorders reduce the fermion conductivity $\sigma$. Thus, even if
the charged fixed point can be stable against weak disorders in
the case of noncompact U(1) gauge fields, the fixed point can be
unstable against instanton excitations owing to reduction of the
conductivity. If so, the fermions would be confined owing to the
presence of disorders. This may be experimentally verified. If
nonmagnetic impurities like $Zn$ are doped in the strange metal
phase of high $T_c$ cuprates\cite{Polchinski,Altshuler,Nagaosa1},
in the quantum critical regime of Kondo
systems\cite{Senthil_Kondo,Kim_Kondo1,Kim_Kondo2}, or in the spin
liquid Mott insulator of geometrically frustrated quantum
antiferromagnetism\cite{Motrunich_SL,SS_SL}, the confinement of
fermions can break quantum criticality, detected in measurements
of conductivity or magnetic susceptibility. This important issue
should be addressed in more quantitative level near future.

\section{Summary}

In the present paper we investigated deconfinement of
non-relativistic fermions near a Fermi surface. The main findings
are the existence of the charged critical point and its stability
against instanton excitations. This leads us to the critical field
theory Eq. (13), where the critical coupling constant between the
fermions and noncompact U(1) gauge fields is given by
$e_{c}/\sqrt{\sigma}$. This coupling strength makes it possible to
calculate correlation functions in the $1/\sigma$ expansion at the
charged critical point.

\section{Acknowledgement}

K.-S. Kim thanks profs. A. Paramekanti and L. Yu for introducing
the present problem in the $9th$ APCTP winter workshop on strongly
correlated electron systems. K.-S. Kim especially thanks prof.
Kim, Yong Baek for his critical reading of this manuscript and
helpful discussions on the present subject.

\appendix

\section{}

In appendix A we sketch the derivation of a fermion polarization
function Eq. (7). We rewrite the current-current correlation
function \bqa && \Pi_{ij}(x, \tau) = -
\langle{T}_{\tau}[J_{Fi}(x,\tau)J_{Fj}(0,0) -
\delta_{ij}\rho_{F}\delta(x)\delta(\tau)]\rangle . \nn \eqa Here
$\vec{J}_{F} =
i\chi_{\alpha}^{\dagger}(\vec{\nabla}/2m)\chi_{\alpha} + h. c.$ is
the fermion current operator and $\rho_{F} =
\chi_{\alpha}^{\dagger}\chi_{\alpha}$, its density operator. The
fermion current operator is reduced to $\vec{J}_{F} =
\vec{v}_{F}\chi_{\alpha}^{\dagger}\chi_{\alpha}$ with $\vec{v}_{F}
= \vec{k}_{F}/m$ at the Fermi energy, consistent with the
expression in section (III-B). Inserting these operators into Eq.
(A1) and performing some algebra, we obtain the following
expression in energy-momentum space \bqa && \Pi_{ij}(q, i\omega)
=- \int\frac{d^2k}{(2\pi)^{2}}\frac{1}{m^2}(k_{i}+
\frac{q_{i}}{2})(k_{j}+\frac{q_{j}}{2})\nn&&~~~~~~~~~~~~~
\frac{1}{\beta}\sum_{\nu}{G}_{0}(k,i\nu)G_{0}(k+q,i\nu+i\omega) ,
\eqa where $G_{0}(k,i\nu) = [i\nu - \epsilon_{k}]^{-1}$ is a
fermion propagator with its bare dispersion $\epsilon_{k} =
k^2/2m$. Here the spin index $\alpha$ is not taken into account.
Performing the sum of Matsubara frequencies of the fermions, we
obtain \bqa && \Pi_{ij}(q, i\omega) =
-\int\frac{d^2k}{(2\pi)^{2}}\frac{1}{m^2}(k_{i}+
\frac{q_{i}}{2})(k_{j}+\frac{q_{j}}{2})
\nn&&~~~~~~~~~~~~~~~~~~~\frac{f(\epsilon_{k}) -
f(\epsilon_{k+q})}{i\omega - [\epsilon_{k+q} - \epsilon_{k}]} ,
\eqa where $f(\epsilon_{k})$ is the Fermi-Dirac distribution
function. Shifting ${\vec k} \rightarrow {\vec k} - {\vec q}/2$,
expanding ${\vec q}$ for $|{\vec q}| << k_{F}$, and using
$\epsilon_{k\pm{q}/2} \approx \epsilon_{k} \pm
\frac{\vec{k}\cdot\vec{q}}{2m}$, one can find \bqa &&
f(\epsilon_{k+q/2}) - f(\epsilon_{k-q/2}) \approx
\frac{\partial{f}(\epsilon_{k})}{\partial\epsilon_{k}}\frac{\vec{k}\cdot\vec{q}}{m}
. \eqa At zero temperature one obtains \bqa &&
\frac{\partial{f}(\epsilon_{k})}{\partial\epsilon_{k}} = -
\delta(\epsilon_{k} - \epsilon_{F}) = - \delta\Bigl(\frac{k^2 -
k_{F}^{2}}{2m}\Bigr) \approx - \frac{m}{k_{F}}\delta(k - k_{F}) .
\nn \eqa Inserting Eq. (A4) and Eq. (A5) into Eq. (A3), Eq. (A3)
reads \bqa && \Pi_{ij}(q, i\omega) \nn && =
-\frac{1}{m^2}\int\frac{d^2k}{(2\pi)^{2}}[k_{i}k_{j}]
\frac{{\vec{k}\cdot\vec{q}}/{m}}{i\omega -
{\vec{k}\cdot\vec{q}}/{m}}\frac{m}{k_{F}}\delta(k - k_{F}) . \eqa
Performing the momentum integration with care of $k_{i}k_{j}$ and
${\vec k}\cdot{\vec q}$, one obtains the following expression for
the transverse current-current correlation
function\cite{Polchinski,Altshuler,Gan,Nagaosa1,Tsvelik} \bqa &&
\Pi_{ij}(q, i\omega) = \Bigl(\delta_{ij} -
\frac{q_{i}q_{j}}{q^{2}}\Bigr)\Pi(q, i\omega) , \nn && \Pi(q,
i\omega) = \sigma|\omega| + \chi{q}^{2} , \eqa where $\sigma$ and
$\chi$ are the conductivity and diamagnetic susceptibility. This
form is quite reasonable because in the $q \rightarrow 0$ limit
the conductivity is reduced to $\sigma = (1/i\omega)\Pi(\omega +
i\epsilon)$ with the Wick rotation $i\omega \rightarrow \omega +
i\epsilon$, and in the $\omega \rightarrow 0$ limit only the
diamagnetic contribution proportional to $q^2$ survives, both of
which are well known. Furthermore, this expression shows that in
the $q, \omega \rightarrow 0$ limit the paramagnetic contribution
(the first term in Eq. (A1)) cancels the diamagnetic one (the
second term in Eq. (A1)) exactly in a normal Fermi
liquid\cite{Nagaosa1}. In this respect the present U(1) gauge
action may be applied to various shapes of Fermi surface. This
statement is justified by the fact that the expression Eq. (A7)
can be derived from the Maxwell equation, as well shown in page
$113$ of Ref. \cite{Tsvelik}. In a free fermion gas the
conductivity is given by $\sigma \sim q^{-1}$, resulting in the
familiar Landau damping term. However, in this paper we consider
the transport time $\tau_{tr}$ due to scattering mechanism such as
disorder. In the case of $q < (v_{F}\tau_{tr})^{-1}$ the
conductivity is given by $\sigma \sim \tau_{tr}$, corresponding to
the Ohm's law\cite{Nagaosa1}. In this paper we consider the Ohmic
dissipation instead of the familiar Landau damping. The
diamagnetic susceptibility is given by $\chi \sim m^{-1}$ from
$\partial^{2}S/\partial{a_{i}}^{2}$, where $S$ is the action
defined in Eq. (5). For more details, see Refs.
\cite{Nagaosa1,Tsvelik}.


\begin{thebibliography}{9}
\bibitem{Herbut_RG} In Ref. \cite{Herbut} the following
non-relativistic sine-Gordon action is investigated \bqa && S_{sG}
= \int\frac{d^2kd\omega}{(2\pi)^{3}}
\frac{1}{2}\Bigl(\frac{|\omega|k^{2}}{l} + a_{k}k^{2} +
a_{\omega}\omega^{2}\Bigr)|\varphi(k,\omega)|^{2} \nn && -
\int{dx^{2}}{d\tau} y_{m}\cos\varphi(x,\tau) . \nonumber \eqa Here
$l$ and $a_{k}$ correspond to $\sigma$ and $1/g$ in the present
consideration, respectively. On the other hand, $a_{\omega}$ is
newly introduced in Ref. \cite{Herbut}. Based on this sine-Gordon
theory the authors derived the following renormalization group
equations \bqa && \frac{d{\hat a}_{k}}{d\ln{b}} = \frac{{\hat
y}^{2}_{m}}{2\pi{\hat a}_{k}} , \nn && \frac{d{\hat
a}_{\omega}}{d\ln{b}} = 2{\hat a}_{\omega} + \frac{{\hat
y}^{2}_{m}}{4{\hat a}_{k}} , \nn && \frac{d{\hat y}_{m}}{d\ln{b}}
= \Bigl(2 - \frac{1}{8\pi\sqrt{{\hat a}_{k}{\hat a}_{\omega} -
(1/2l)^{2}}} \Bigr){\hat y}_{m} \nonumber \eqa with the redefined
variables, ${\hat a}_{k} = a_{k}/\Omega$, ${\hat a}_{\omega} =
a_{\omega}\Omega/\Lambda^{2}$, and ${\hat y}_{m} =
y/\Lambda^{2}\Omega$, where $\Omega$ and $\Lambda$ are energy and
momentum cut-offs, respectively. $b$ is a scaling parameter.
Notice that these renormalization group equations result in ${\hat
y}_{m} \rightarrow \infty$ owing to the newly introduced parameter
${\hat a}_{\omega}$, indicating condensation of magnetic monopoles
(instantons). The main problem in these renormalization group
equations is that even in the absence of instanton excitations,
i.e., ${\hat y}_{m} \equiv 0$ the remaining gaussian action in the
above sine-Gordon action $S_{sG}$ is not a critical theory owing
to the ${\hat a}_{\omega}$ parameter. See the second
renormalization group equation for ${\hat a}_{\omega}$ showing the
relevance of ${\hat a}_{\omega}$ in the case of ${\hat y}_{m}
\equiv 0$. This is in contrast to the previous
studies\cite{Polchinski,Altshuler,Gan} and the present result
claiming the existence of a scale invariant critical point in the
case of noncompact U(1) gauge fields. In this respect we set the
parameter ${\hat a}_{\omega}$ to be identically zero. In this case
the resulting renormalization group equations are given
by\cite{Herbut} \bqa && \frac{d{\hat a}_{k}}{d\ln{b}} =
\frac{{\hat y}^{2}_{m}}{2\pi{\hat a}_{k}} , \nn && \frac{d{\hat
y}_{m}}{d\ln{b}} = \Bigl[2 - \frac{l}{4\pi^{2}}\ln\Bigl(1 +
\frac{1}{l{\hat a}_{k}}\Bigr) \Bigr]{\hat y}_{m} . \nonumber \eqa
These renormalization group equations have a fixed line, ${\hat
y}_{m} = 0$ and ${\hat a}_{k} \leq \Bigl[l\Bigl(e^{8\pi^2/l} -
1\Bigr)\Bigr]^{-1}$ as the $KT$ transition\cite{Kleinert1}. This
deconfinement transition differs from our scenario in the respect
that the deconfinement in the present paper can occur at the
charged critical point instead of a critical line.
\end{thebibliography}
\end{document}